# Mesoporous alumina- and silica-based crystalline nanocomposites with tailored anisotropy: methodology, structure and properties

A.V. Kityk[1]*, A. Andrushchak[2], Ya. Shchur[3], V.T. Adamiv[4], O. Yaremko[2], M. Lelonek[5], S.A. Vitusevich[6], O. Kityk[7], R. Wielgosz[7], W. Piecek[8], M. Busch[9], K. Sentker[9], P. Huber[9]

[1]Faculty of Electrical Engineering, Czestochowa University of Technology, Al. Armii Krajowej 17, 42-200 Czestochowa, Poland
[2]Institute of Telecommunications, Radio Electronics and Electronic Technics, Lviv Polytechnic National University, 12 S. Bandery str., 79013 Lviv, Ukraine
[3]Institute for Condensed Matter Physics, 1 Svientsitskii str., 79011 Lviv, Ukraine
[4]O.G. Vlokh Institute of Physical Optics, 23 Dragomanova str., 79005 Lviv, Ukraine
[5]SmartMembranes GmbH, Heinrich-Damerow Str. 4, D-06120 Halle, Germany
[6]Institute of Bioelectronics (ICS-8), Forschungszentrum Jülich, D-52425 Jülich, Germany
[7]Energia Oze Sp. z o.o., ul. Częstochowska 7, 42-274 Konopiska, Poland
[8]Military University of Technology, 00-908 Warsaw, Poland
[9]Institute of Materials Physics and Technology, Hamburg University of Technology, Eissendorferstr. 42, D-21073 Hamburg, Germany
*Tel: (4834) 3250 815, Fax: (4834) 3250 823, e-mail: kityk@ap.univie.ac.at

**ABSTRACT**

We present several recently synthesized nanocomposites consisting of liquid crystals as well as an organic molecular crystal embedded into the nanochannels of mesoporous alumina and silica. As liquid-crystalline mesogens achiral, nematogen and chiral cholesteric guest molecules infiltrated into nanochannels by spontaneous imbibition were chosen. The molecular ordering inside the nanochannels, which can be tailored by modifying the surface anchoring, was characterized by optical polarimetry (linear and/or circular birefringence) in combination with X-ray diffraction. For the synthesis of the solid crystalline nanocomposites ferroelectric triglycine sulfate (TGS) nanocrystals were deposited into the nanochannels by slow evaporation of saturated water solutions imbibed into the porous hosts. Their textural and physicochemical properties were explored by x-ray diffraction, scanning electron microscopy and dielectric techniques.

**Keywords**: nanocomposites, nanocrystals, liquid-crystalline nanocomposites, optical anisotropy, linear and circular birefringence, dielectric properties

## 1. INTRODUCTION

Mesoporous alumina ($p$Al$_2$O$_3$, anodized aluminium oxide - AAO) and silica ($p$SiO$_2$) represent nanoporous host templates for a variety of nanoscale technologies. In particular liquid- and solid crystalline-filled templates of these materials appear as prominent candidates for prospective optoelectronic and photonic applications. The unidirectional tubular structure of such matrices results in a macroscopic anisotropy of many materials properties. Moreover, it is characterized by a remarkable flexibility and modularity: it is possible to tailor the properties of these nanocomposites according to specific needs by varying geometrical characteristics of the pore network (channel diameters and/or porosity), by choosing suitable guest components, by modifying the interface anchoring or nanocrystals growth conditions [1-3]. Here, we present several recently synthesized nanocomposites with liquid-crystalline and a solid-crystalline filling embedded into the nanochannels of mesoporous alumina and silica.

## 2. EXPERIMENTAL RESULTS AND DISCUSSION

Mesoporous AAO membranes were fabricated by electrochemical two stage etching using oxalic acid (0.3 M water solution) as electrolyte. Nanoporous silica, $p$SiO$_2$, was synthesized by oxidation of mesoporous silicon obtained by electrochemical etching of $p$-doped $\langle 100 \rangle$ Si-wafer in an electrolyte composed of a mixture of HF and ethanol with a volume ratio 2:3. Liquid-crystal based nanocomposites were obtained by embedding liquid crystals into tubular pore networks of mesoporous alumina or silica membranes by capillarity-driven imbibition. The surface anchoring at the pore walls is crucial for the molecular ordering in pore space and thus defines the optical anisotropy of liquid crystal nanocomposites. Untreated (native) surfaces of alumina or silica pore walls are hydrophilic. They can be made hydrophobic surface by silanization resulting in changed molecular anchoring at the pore walls. Similarly, polymer coatings, SE130 or SE1211, enhance tangential or normal molecular anchoring. The anisotropy of nanocomposite materials are characterized with optical polarimetry by measuring linear birefringence, associated with an optical retardation $R$, and/or circular birefringence, associated with optical activity $\rho$. Temperature changes of these quantities (see Fig.1, left) are related with phase transformations and thus collective orientational and/or translational molecular rearrangements in pore space, see

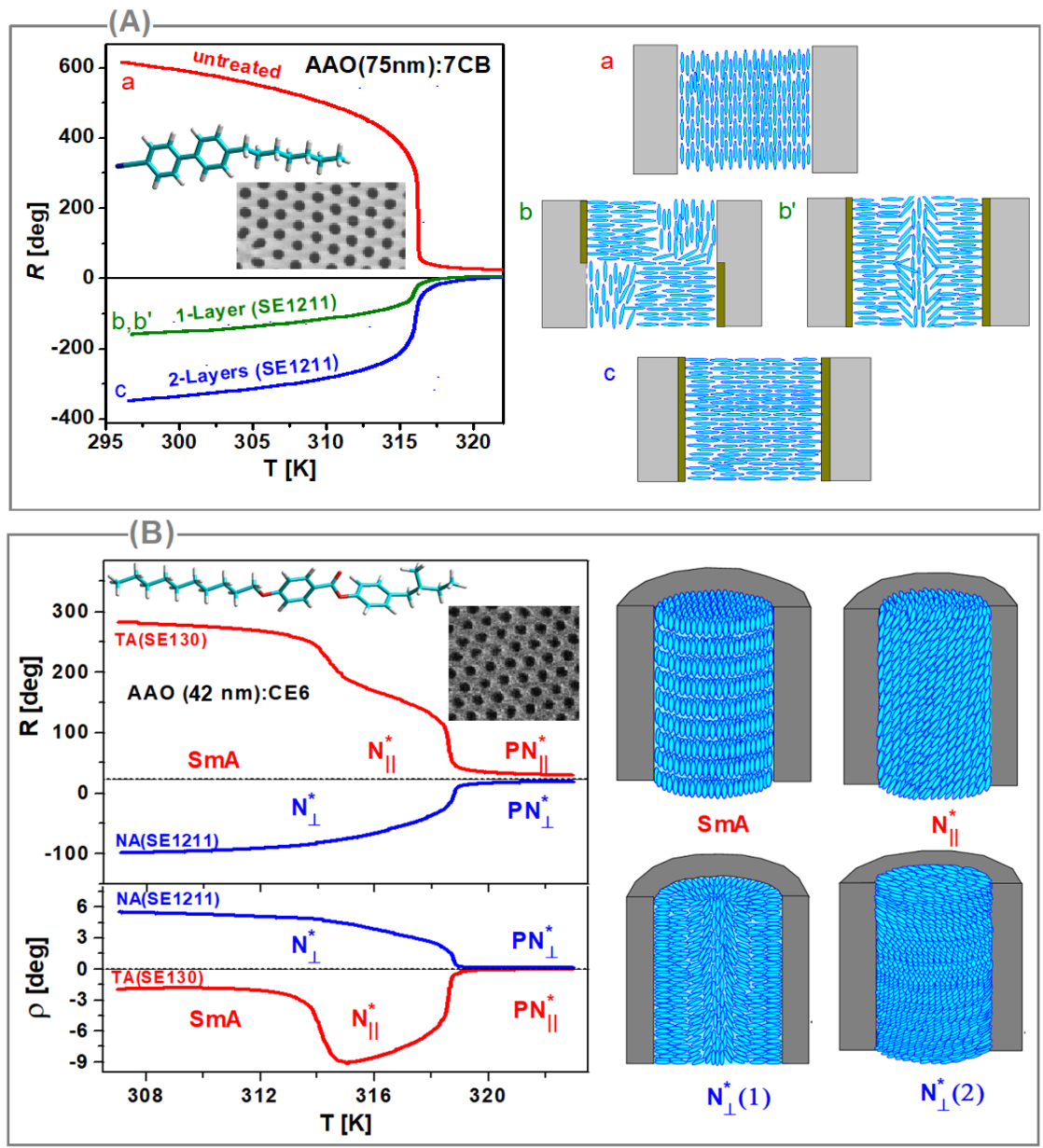

*Figure 1. Examples of the liquid-crystal based nanoporous composites with inorganic host AAO matrix of different pore sizes (membrane thickness h=0.1 mm) and guest liquid crystal molecules of rod-like cyanobiphenyl nematic 7CB (section A) and chiral nematic (cholesteric) CE6 (section B). Molecular ordering (see sketches on the right) and related optical anisotropy of the nanocomposites, i.e. their linear birefringence (optical retardation $R \sim \Delta n$, incident angle $36^o$) and/or circular birefringence (optical rotation $\rho$) (left) can be tailored by modifying the surface anchoring. Polymer pore-surface coverages, e.g. SE130 or SE1211 enhance tangential or normal anchoring, respectively.*

Fig. 1 (right). Untreated pore walls, likewise the polymer SE130 coated ones, exhibit tangential anchoring resulting in a positive birefringence as demonstrated in Fig.1 for confined achiral nematic (section A) and chiral (cholesteric) nematic or SmA (section B) phases. The sign of the birefringence can be turned to negative by enhancing hydrophobicity of the nanochannels, particularly by silanization or by using appropriate sub-nanometric polymer coverages, such as e.g. SE1211.

This pore-surface grafting leads in the case of confined achiral nematics to either escaped radial or polar molecular configurations, see sketches (b') and (c), respectively, in Fig.1 (A), right panel. Amazingly, a two-layer polymer coverage enhances this transition. A single layer polymer coverage, apparently, is not homogeneous and exhibits local islands (surface heterogeneities) as it is schematically illustrated in sketch (b) of Fig. 1(A), right panel. Accordingly, a coexistence of nanodomains with positive and negative anisotropy is expected causing, effectively, a weak negative birefringence, see Fig. 1(A), left panel.

A cholesteric liquid crystal (CE6) embedded into the polymer treated nanochannels with tangential anchoring (SE130) reorganizes at cooling from an isotropic phase to a so-called cholesteric double-twist structure and then to a smectic A structure, see sketches in Fig.1 (section B, right) labelled as $N_\parallel^*$ and SmA, respectively [1]. Both phases are characterized by a positive birefringence. The accompanying optical activity $\rho$ in the confined cholesteric phase indicates a formation a long-range helical structure aligned along the channel axis. Normal anchoring, in contrast, results in a negative optical birefringence and opposite rotation of the light polarization. The smectic A phase, observed in bulk CE6, appears to be completely suppressed here [1]. Negative optical birefringence ($\Delta n = n_e - n_o < 0$) combined with considerable optical activity suggest an either escaped radial $N_\perp^*(1)$ or simple cholesteric twist structure $N_\perp^*(2)$, as sketched in Fig.1(B), right.

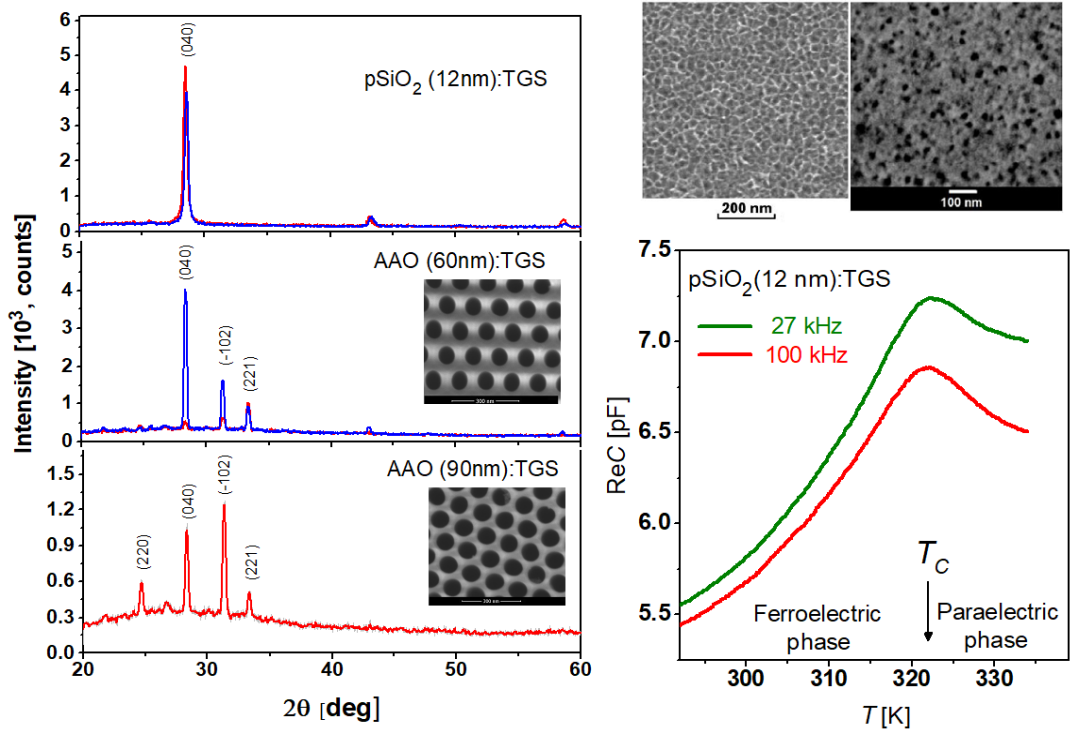

*Figure 2. Left: X-ray diffraction patterns (XRD) of the silica (pSiO$_2$, d=12 nm) and alumina (AAO, d=60 nm and 90 nm) based nanocomposites with embedded TGS nanocrystals. The inserts show SEM images of the host AAO membranes. Blue and red online colours correspond to XRDs with reflections recorded from different sides of the nanocomposite membranes. Right: SEM images of mesoporous silica templates pSiO$_2$ (mean pore diameter 12 nm). Temperature dependences of the real part of the complex capacitance, ReC ~ ε, of pSiO$_2$(12 nm):TGS nanocomposite (right) and SEM images of relevant porous silica templates.*

Nanocomposites with embedded solid crystalline materials, e.g. nanocrystals, represent an alternative approach to achieve materials with tailorable physical properties. We followed this strategy by the deposition of guest ferroelectric crystals TGS into the nanochannels via slow evaporation of TGS-saturated water solutions imbibed into the pore network. The textural properties of the resulting soft-hard hybrid materials, which turn out to be strongly dependent on the host/guest components and channels size, we explore by X-ray and SEM techniques. In Fig.2 (left) we present x-ray diffractograms (XRDs) of several nanocomposites with ferroelectric nanocrystals TGS embedded into nanochannels of mesoporous silica (pSiO$_2$(12 nm):TGS) and alumina (AAO(60 nm):TGS and AAO(90 nm):TGS) membranes. The inserts show SEM images of the host AAO matrices characterized by highly ordered tubular pore network arranged hexagonally as obtained by the double-stage electrochemical etching method. Nanocomposites with small channel sizes, e.g. pSiO$_2$(12 nm):TGS, are strongly dominated by guest TGS nanocrystals with a strong texture. As evidenced by the dominance of the (040) Bragg reflection in our scattering experiment, where the wave-vector transfer is parallel to the pore axis, the polar axis of the nanoconfined monoclinic TGS crystals is oriented parallel to the long channel axis. The solid crystal nanocomposites with larger channel diameters, on the other hand, such as e.g. AAO(60 nm):TGS or AAO(90 nm):TGS, are characterized by a more random orientation of the nanocrystals in pore space. The XRD patterns exhibit a more powder-like Bragg peak distribution, see Fig.2. Moreover, diffractograms recorded in reflection geometry from both sides of these nanocomposite membranes exhibit different Bragg reflection patterns indicating an inhomogeneous nanocrystal content distribution and/or their orientation along the nanochannels.

The nanocomposites with the most homogeneous content and uniform orientation of TGS nanocrystals with ferroelectric axes parallel to the long channel axis, particularly $p$SiO$_2$(12 nm):TGS, may be of particular practical interest. One must emphasize that bulk TGS at room temperature is a ferroelectric crystal and along the polar axis it exhibits piezoelectricity, piroelectricity and a large dielectric response. The dielectric constant is characterized by a Curie-Weiss temperature behaviour, i.e. it anomalously diverges in the vicinity of the Curie point ($T_C$=322 K) [4]. The dielectric behaviour of $p$SiO$_2$(12 nm):TGS nanocomposite, in contrast, exhibits only a weak rising in this temperature region (see Fig.2, right panel) apparently confirming the ferroelectric origin of the embedded nanocrystals. A significant damping of the dielectric anomaly compared to the bulk crystalline state may be explainable by an incomplete filling of the channels with nanocrystals. Therefore, the electric field drops in the large number of voids (series capacitance) resulting in a drastic reduction of the effective dielectric susceptibility.

## 3. CONCLUSIONS

We have presented several examples of nanocomposite materials consisting of host mesoporous tubular alumina or silica membranes with embedded achiral as well as chiral liquid crystals and ferroelectric nanocrystals. In the case of the liquid crystal based nanocomposites the molecular ordering inside the nanochannels can be controlled by the surface anchoring. Thus the optical anisotropy and likewise other physicochemical properties can be tailored. In the developed solid crystalline nanocomposites the guest TGS ferroelectric nanocrystals have been deposited into the nanochannels of mesoporous silica and alumina by slow evaporation of their saturated water solution imbibed by the pore network. Their physicochemical and textural properties, which turn out to be strongly dependent on the host/guest components and channel size, are explored by X-ray, SEM and dielectric spectroscopy techniques. For the future alternative guest materials, particularly, ferroelectric liquid crystals, achiral and chiral discotics and twist-bend nematics as well as other water soluble classic ferroelectric crystals, like e.g. KDP [4,5] or GPI [6] and series of proper and improper incommensurate ferroelectric and ferroelastic crystals of the A$_2$BX$_4$ group [7-9] would be of particular scientific and technological interest.

**ACKNOWLEDGEMENTS**

The presented results are part of a project that has received funding from the European Union's Horizon 2020 research and innovation programme under the Marie Skłodowska-Curie grant agreement No 778156. The support from Ministry of Education and Science of Ukraine (project 0119U002255 "Nanocrystallite") is acknowledged. A.V.K. acknowledges support from resources for science in the years 2018-2022 granted for the realization of international co-financed project Nr W13/H2020/2018 (Dec. MNiSW 3871/H2020/2018/2). Ya.S. was supported by project 0118U003010 of National Academy of Sciences of Ukraine. K.S. and P.H. acknowledge funding by the Deutsche Forschungsgemeinschaft (DFG, German Research Foundation) – Projektnummer 192346071 – SFB 986. W.P. acknowledges support by the Ministry of National Defence Republic of Poland Program- Research Grant MUT Project 13- 995.